\documentclass[11pt]{article}
\usepackage{amsfonts,amssymb,amsthm,graphicx,rotating,appendix,xcolor,helvet,amsmath}
\usepackage[ansinew]{inputenc}
\usepackage[english]{babel}
\usepackage{algorithm}
\floatname{algorithm}{Algorithm}

\newtheorem{thm}{Theorem}[section]

\newtheorem{lem}[thm]{Lemma}
\theoremstyle{definition}

\theoremstyle{remark}

\newtheorem{prop}{Proposition}

\newtheorem{lemproof}{Proof}

\numberwithin{equation}{section}

\begin{document}

\title{ {\bf Exact balanced random imputation for sample survey data}}

\author{Guillaume Chauvet\thanks{ENSAI/IRMAR, Campus de Ker Lann, 35170 Bruz, France;} \and Wilfried Do Paco\thanks{Insee}}

\maketitle

\begin{abstract}
\noindent Surveys usually suffer from non-response, which decreases the effective sample size. Item non-response is typically handled by means of some form of random imputation if we wish to preserve the distribution of the imputed variable. This leads to an increased variability due to the imputation variance, and several approaches have been proposed for reducing this variability. Balanced imputation consists in selecting residuals at random at the imputation stage, in such a way that the imputation variance of the estimated total is eliminated or at least significantly reduced. In this work, we propose an implementation of balanced random imputation which enables to fully eliminate the imputation variance. Following the approach in Cardot et al. (2013), we consider a regularized imputed estimator of a total and of a distribution function, and we prove that they are consistent under the proposed imputation method. Some simulation results support our findings.
\end{abstract}
\vspace*{.30in}

\section{Introduction} \label{sec:1}

\noindent Even if survey staff do their best in order to maximize response, it is unavoidable that surveys will suffer from some degree of non-response. This makes the effective sample size smaller, which results in an increase of the variance of estimators. More importantly, the respondents usually differ from the non-respondents with respect to the study variables. Therefore, unadjusted estimators will tend to be biased. In order to reduce the so-called non-response bias, it is therefore necessary to define estimation procedures accounting for non-response. Survey statisticians usually distinguish unit nonresponse from item nonresponse. The former occurs when all variables are missing for some sampled unit, which may be due to a refusal to participate to the survey, or to the impossibility to contact the sampled unit, for example. The latter occurs when some variables, but not all, are missing for some sampled unit, which may be due to a refusal to answer to certain delicate questions in the survey, or to the length of the questionnaire, for example. Unit non-response is typically accounted for by reweighting estimators. Item non-response is typically handled by means of some form of imputation, which consists in replacing missing values with artificial values in order to reduce the bias and possibly control the variance due to non-response. In this paper, we are interested in imputation procedures to treat item non-response. \\

\noindent Imputation methods may be classified into two broad classes: deterministic and random. Deterministic imputation methods yield a fixed imputed value given the sample. For example, deterministic regression imputation consists in using a regression model to predict the missing value for a non-respondent, making use of auxiliary information available for the whole sample, including non-respondents. Deterministic regression imputation leads to an approximately unbiased estimator of the total if the regression model is correctly specified. However, deterministic imputation tends to distort the distribution of the imputed variable, and some form of random imputation is typically used if we wish to preserve the distribution of the imputed variable. Random imputation methods are closely related to deterministic imputation methods, except that a random term is added to the prediction in order to mimic as closely as possible the relationship between the variable of interest and the explanatory variables. The main drawback of random imputation methods is that they lead to estimators with increased variability due to the imputation variance. In some cases, the contribution of the imputation variance to the global variance may be large, resulting in potentially inefficient estimators. \\

\noindent The data collected for a given survey are typically used to estimate a variety of parameters. The survey is often primarily designed to estimate totals over a given population of interest. For example, this may be the total biomass of living vegetation for a region in case of a forest survey (Gregoire and Valentine, 2008, page 3),
or the total of revenues and expenses inside categories of firms in case of business surveys (Haziza et al., 2016). On the other hand, secondary analysts may be interested in estimating more complex parameters, some of them being directly linked to the population distribution function like the quantiles (Boistard et al., 2016). Therefore, a same variable of interest is habitually used to estimate several parameters. A random imputation method is needed if we wish to preserve the distribution of this variable, but the imputation mechanism should be chosen so that the imputation variance is kept as small as possible for the estimation of the total of the variable. \\

\noindent In the literature, three general approaches for reducing the imputation variance have been considered. The fractional imputation approach consists of replacing each missing value with $M \geq 2$ imputed values selected randomly, and assigning a weight to each imputed value (Kalton and Kish, 1981, 1984; Fay, 1996; Kim and Fuller, 2004; Fuller and Kim, 2005). It can be shown that the imputation variance decreases as $M$ increases. The second approach consists of first imputing the missing values using a standard random imputation method, and then adjusting the imputed values in such a way that the imputation variance is eliminated; see Chen et al. (2000). The third approach that we study consists of selecting residuals at random in such a way that the imputation variance is eliminated (Kalton and Kish, 1981, 1984; Deville, 2006; Chauvet et al., 2011; Hasler and Till\'e, 2014). \\

\noindent In this work, we propose an implementation of balanced random imputation that makes it possible to fully eliminate the imputation variance for the estimation of a total. Also, we propose regularized imputed estimators of a total and of the distribution function, following the approach in Cardot et al. (2013), and we establish their consistency. The paper is organized as follows. Our notations in case of full response are defined in Section \ref{sec:frame}, and the principles of balanced sampling are briefly reminded. The imputation model is presented in Section \ref{sec:imput:mod}, along with the imputed estimators of the total and of the distribution function. The regularized estimator of the model parameter is also introduced. In Section \ref{sec:bal:imput:model}, we describe the proposed exact balanced random imputation method. We give an illustration on a small dataset, and we prove the consistency of the imputed estimator of the total and of the imputed estimator of the distribution function. The results of a simulation study are presented in Section \ref{sec:simu:study}. We draw some conclusions in Section \ref{sec:conc}. The proofs are deferred to the Appendix.

\section{Finite population framework} \label{sec:frame}

\subsection{Notation} \label{ssec:11}

\noindent We consider a finite population $U$ of size $N$ with some variable of interest $y$. We are interested in estimating some finite population parameter such as the total $t_y=\sum_{k \in U} y_k$ or the population distribution function
    \begin{eqnarray} \label{FN:t}
      F_N(t) & = & N^{-1} \sum_{k \in U} 1(y_k \leq t)
    \end{eqnarray}
with $1(\cdot)$ the indicator function. \\

\noindent In order to study the asymptotic properties of the sampling designs and estimators that we treat below, we consider the asymptotic framework of Isaki and Fuller~(1982). We assume that the population $U$ belongs to a nested sequence $\{U_t\}$ of finite populations with increasing sizes $N_t$, and that the population vector of values $y_{Ut}=(y_{1t},\ldots,y_{Nt})^{\top}$ belongs to a sequence $\{y_{Ut}\}$ of $N_t$-vectors. For simplicity, the index $t$ will be suppressed in what follows and all limiting processes will be taken as $t \to \infty$. \\

\noindent A random sample $S$ is selected in $U$ by means of some sampling design $p(\cdot)$, which is a probability distribution defined over the subsets of the population $U$. That is, we have
    \begin{eqnarray} \label{samp:des}
    p(s) \geq 0 \textrm{ for any } s \subset U \textrm{ and } \sum_{s \subset U} p(s) = 1.
    \end{eqnarray}
We assume that the sampling design is of fixed size $n$, which means that a subset $s$ has a probability of selection equal to zero if this subset is not of size $n$. We note $I_k$ for the sample membership indicator, equal to $1$ if the unit $k$ is selected in the sample $S$ and to $0$ otherwise. We note $I_U=(I_1,\ldots,I_N)^{\top}$ for the vector of sample indicators. Since the sampling design is of fixed size $n$, we have
    \begin{eqnarray} \label{fixed:size}
    \sum_{k \in U} I_k & = & n.
    \end{eqnarray}

\subsection{Inclusion probabilities} \label{ssec:12}

\noindent The probability for unit $k$ to be included in the sample is denoted as $\pi_k$. We note $\pi_U=(\pi_1,\ldots,\pi_N)^{\top}$ for the vector of inclusion probabilities. All the inclusion probabilities are assumed to be non-negative, i.e. there is no coverage bias in the population. Since $\pi_k=E_p(I_k)$, with $E_{p}$ the expectation with respect to the sampling design $p(\cdot)$, we obtain from equation (\ref{fixed:size}) that
    \begin{eqnarray} \label{sum:probas}
    \sum_{k \in U} \pi_k & = & n.
    \end{eqnarray}
We also denote by $\pi_{kl}$ the probability that units $k$ and $l$ are selected jointly in the sample. \\

\noindent In case of equal inclusion probabilities, we have $\pi_k=n/N$ for any unit $k \in U$. This occurs for example if the sample is selected by means of simple random sampling without replacement. Another customary choice consists in using inclusion probabilities proportional to some auxiliary non-negative variable $z_{1k}$, known for any unit $k \in U$. This leads to a so-called probability proportional to size ($\pi$-ps) sampling design, which is used in some business surveys (e.g., Ohlsson, 1998). In such case, we obtain from (\ref{sum:probas}) that
    \begin{eqnarray} \label{pi:ps}
      \pi_k & = & n \frac{z_{1k}}{\sum_{l \in U} z_{1l}}.
    \end{eqnarray}
If some units exhibit a large value for the auxiliary variable $z_{1k}$, equation (\ref{pi:ps}) may lead to inclusion probabilities greater than $1$. In this case, these inclusion probabilities are set to $1$, which means that the corresponding units are selected in the sample with certainty, and the inclusion probabilities for the remaining units are computed by means of equation (\ref{pi:ps}) restricted to the remaining units (see Till\'e, 2006, page 18). \\

\noindent In a situation of full response, a design-unbiased estimator for $t_y$ is the Horvitz-Thompson estimator
    \begin{eqnarray} \label{est:ht}
      \hat{t}_{y\pi} & = & \sum_{k \in U} d_k I_k y_k = \sum_{k \in S} d_k y_k
    \end{eqnarray}
with $d_k=\pi_k^{-1}$ the sampling weight, and an approximately unbiased estimator for $F_N(t)$ is
    \begin{eqnarray} \label{est:ht:df}
      \hat{F}_{N}(t) = \frac{1}{\hat{N}} \sum_{k \in S} d_k 1(y_k \leq t) & \textrm{ with } & \hat{N}=\sum_{k \in S} d_k.
    \end{eqnarray}

\subsection{Balanced sampling} \label{ssec:13}

\noindent Suppose that a $q$-vector $x_k$ of auxiliary variables is known at the design stage for any unit $k \in U$. A sampling design $p(\cdot)$ is said to be balanced on $x_k$ if the vector $I_U$ of sample indicators is such that
    \begin{eqnarray} \label{bal:eq}
      \sum_{k \in U} \frac{x_k}{\pi_k} I_k & = & \sum_{k \in U} x_k.
    \end{eqnarray}
In other words, the sampling design is balanced on $x_k$ if for any possible sample, the Horvitz-Thompson estimator of the total of the auxiliary variables exactly matches the true total. \\

\noindent Deville and Till\'e~(2004) introduced a sampling design called the cube method, which enables to select balanced samples, or approximately balanced samples if an exact balancing is not feasible. The cube method proceeds through a random walk from the vector $\pi_U$ of inclusion probabilities to the vector $I_U$ of sample indicators. This random walk proceeds in two steps. At the end of the first one called the flight phase (see Appendix \ref{app:flight}), we obtain a random vector $\tilde{I}_U=(\tilde{I}_1,\ldots,\tilde{I}_N)^{\top}$ such that
    \begin{eqnarray} \label{bal:eq:2}
      \sum_{k \in U} \frac{x_k}{\pi_k} \tilde{I}_k & = & \sum_{k \in U} x_k,
    \end{eqnarray}
and such that $\tilde{I}_k=0$ if the unit $k$ is definitely rejected from the sample, $\tilde{I}_k=1$ if the unit $k$ is definitely selected in the sample, and $0<\tilde{I}_k<1$ if the decision for unit $k$ remains pending. From equation (\ref{bal:eq:2}), the balancing is exactly respected at the end of the flight phase, but we do not obtain a sample per se since the decision remains pending for some units. From Proposition 1 in Deville and Till\'e~(2004), it can be shown that the number of such units is no greater than $q$, the number of auxiliary variables. A second step called the landing phase is then applied on the set of remaining units, in order to end the sampling while ensuring that the balancing equations (\ref{bal:eq}) remain approximately satisfied. This leads to the vector of sample indicators $I_U$.

\section{Imputed estimators} \label{sec:imput:mod}

\noindent In a situation of item non-response, the variable $y$ is observed for a subsample of units only. We note $r_k$ for a response indicator for unit $k$, and $\phi_k$ for the response probability of unit $k$. We note $n_r$ the number of responding units, and $n_m$ the number of missing units. We assume that the units respond independently. In case of simple imputation, an artificial value $y_k^*$ is used to replace the missing $y_k$ and leads to the imputed version of the HT-estimator
    \begin{eqnarray} \label{est:ht:imp}
      \hat{t}_{yI} & = & \sum_{k \in S} d_k r_k y_k + \sum_{k \in S} d_k (1-r_k) y_k^*,
    \end{eqnarray}
and to the imputed version of the estimated distribution function
    \begin{eqnarray} \label{est:ht:df:imp}
      \hat{F}_I(t) & = & \frac{1}{\hat{N}} \left\{\sum_{k \in S} d_k r_k 1(y_k \leq t) + \sum_{k \in S} d_k (1-r_k) 1(y_k^* \leq t)\right\}.
    \end{eqnarray}

\subsection{Imputation model} \label{ssec:31}

\noindent Many imputation methods used in practice can be motivated by the general model
    \begin{equation} \label{imp:mod}
      m:~y_k=f(z_k,\beta)+v_k^{1/2} \epsilon_k,
    \end{equation}
where $f(\cdot)$ is a given function, $z_k$ is a $K$-vector of auxiliary variables available at the imputation stage for all $k \in S$, $\beta$ is a $K$-vector of unknown parameters and $v_k$ is a known constant. The $\epsilon_k$ are assumed to be independent and identically distributed random variables with mean $0$ and variance $\sigma^2$, with a common distribution function denoted as $F_{\epsilon}(\cdot)$ and where $\sigma$ is an unknown parameter. The model (\ref{imp:mod}) is often called an imputation model (e.g., S\"arndal, 1992; Chauvet et al, 2011). \\

\noindent In practice, most imputation techniques which are used in surveys are motivated by a particular case of the imputation model $m$ when $f(z_k,\beta)=z_k^{\top}\beta$. This is true for mean imputation, where a missing value is replaced by the mean of respondents; for hot-deck imputation, where a missing value is replaced by randomly selecting an observed value among the respondents; for regression imputation, where a missing value is replaced by a prediction by regression, to which a random noise is added in case of random regression imputation (see Haziza, 2009). In order to simplify the presentation, we therefore focus on the linear case $f(z_k,\beta)=z_k^{\top}\beta$ in the remainder of the paper, which leads to the so-called regression imputation model
    \begin{equation} \label{imp:mod:2}
      m:~y_k=z_k^{\top}\beta+v_k^{1/2} \epsilon_k.
    \end{equation}
A particular case of model (\ref{imp:mod:2}) called the ratio imputation model is presented in Section \ref{ssec:illust} for illustration. \\

\noindent In this paper, inference will be made with respect to the joint distribution induced by the imputation model, the sampling design and the non-response mechanism, which is known as the Imputation Model approach (IM). We assume that the sampling design is non-informative (S\"arndal et al, 1992, p. 33; Pfeffermann, 2009), namely that the vector of sample membership indicators $I_U \equiv (I_1,\ldots,I_N)^{\top}$ is independent of $\epsilon_U \equiv (\epsilon_1,\ldots,\epsilon_N)^{\top}$, conditionally on a set of design variables $x_U \equiv (x_1,\ldots,x_N)^{\top}$. We do not need an explicit modeling of the non-response mechanism, unlike the Non-response Model approach (see Haziza, 2009). However, the data are assumed to be missing at random (see Rubin, 1976, 1983) in the sense that the vector of response indicators $r_U \equiv (r_1,\ldots,r_N)^{\top}$ is related to a set of auxiliary variables $z_U \equiv (z_1,\ldots,z_N)^{\top}$ known for any unit $k$ in $S$, but the vector $r_U$ is independent of the vector $y_U$, conditionally on $z_U$. 

\subsection{Imputation mechanism} \label{ssec:32}

\noindent Mimicking the imputation model (\ref{imp:mod:2}), the imputed value is
    \begin{equation} \label{val:imp}
      y_k^*=z_k^{\top} \hat{B}+v_k^{1/2} \epsilon_k^*,
    \end{equation}
with $\hat{B}$ some estimator of $\beta$. Using $\epsilon_k^*=0$ in (\ref{val:imp}), we obtain deterministic regression imputation which leads to an approximately unbiased estimation for the total $t_y$  but not for the distribution function $F_N(\cdot)$. Therefore, we focus in the rest of the paper on random regression imputation, where the imputed values in (\ref{val:imp}) are obtained by generating the random residuals $\epsilon_k^*$ randomly. For each unit $k$ for which $y_k$ is missing, we select a donor, which is a responding unit for which the value of the variable of interest is used to fill-in the missing value for unit $k$. More precisely, the random residuals $\epsilon_k^*$ are selected from the set of observed residuals
    \begin{eqnarray} \label{e_l}
      E_r=\{e_l;~r_l=1\} & \textrm{ where } & e_l=\frac{y_l - z_l^{\top} \hat{B}}{v_l^{1/2}}.
    \end{eqnarray}
The residual $e_l$ is attributed to the non-respondent $k$ with the probability
    \begin{eqnarray} \label{pr:epsk:star}
      Pr(\epsilon_k^*=e_l)=\tilde{\omega}_l & \textrm{ where } & \tilde{\omega}_l = \frac{\omega_k}{\sum_{l \in S} \omega_l r_l},
    \end{eqnarray}
where $\omega_l$ is an imputation weight attached to unit $l$. We assume that these imputation weights do not depend on $\epsilon_U$, $I_U$ or $r_U$. Alternatively, the residuals $\epsilon_k^*$ could be generated from a given parametric distribution. \\

\noindent A possible estimator for the unknown parameter $\beta$ is
    \begin{eqnarray} \label{Br:hat}
      \hat{B}_r=\hat{G}_r^{-1} \left(\frac{1}{N} \sum_{k \in S} r_k \omega_k v_k^{-1} z_k y_k \right) & \textrm{ with } & \hat{G}_r=\frac{1}{N} \sum_{k \in S} r_k \omega_k v_k^{-1} z_k z_k^{\top}.
    \end{eqnarray}
Since a matrix $\hat{G}_r$ close to singularity can lead to unstable estimators, we follow the approach proposed in Cardot et al. (2013) and we introduce a regularized version of $\hat{B}_r$. We first write
    \begin{eqnarray}
      \hat{G}_r = \sum_{j=1}^p \eta_{jr} u_{jr} u_{jr}^{\top},
    \end{eqnarray}
where $\eta_{1r} \geq \ldots \geq \eta_{pr}$ are the non-negative eigenvalues of $\hat{G}_r$, with $u_{1r},\ldots,u_{pr}$ the corresponding orthonormal eigenvectors. For a given $a>0$, the regularized version of $\hat{G}_r$ as defined in Cardot et al. (2013) is then
    \begin{eqnarray}
      \hat{G}_{ar} = \sum_{j=1}^p \max(\eta_{jr},a) u_{jr} u_{jr}^{\top},
    \end{eqnarray}
which is an invertible matrix with
    \begin{eqnarray} \label{norm:Gar:inv}
      \| \hat{G}_{ar}^{-1} \| & \leq & a^{-1},
    \end{eqnarray}
where $\|\cdot\|$ stands for the spectral norm. This leads to the regularized estimator of the parameter $\beta$
    \begin{eqnarray} \label{Br:hat}
      \hat{B}_{ar}=\hat{G}_{ar}^{-1} \left(\frac{1}{N} \sum_{k \in S} r_k \omega_k v_k^{-1} z_k y_k \right).
    \end{eqnarray}
In the rest of the paper, we use this regularized estimator in (\ref{val:imp}) to generate the imputed values, and in (\ref{e_l}) to define the observed residuals. 

\subsection{Consistency of the regularized estimator} \label{ssec:33}

\noindent In order to study the asymptotic properties of the estimators that we treat below, we consider the following regularity assumptions:
    \begin{itemize}
      \item[] H1: There exists some constant $C_1,C_2>0$ such that $C_1 \leq N n^{-1} \pi_k \leq C_2$ for any $k \in U$.
      \item[] H2: There exists some constant $C_3$ such that $\sup_{k \neq l \in U} \left(n \left|1-\frac{\pi_{kl}}{\pi_k\pi_l}\right|\right) \leq C_3$.
      \item[] H3: There exists some constant $C_4>0$ such that $C_4 \leq \min_{k \in U} \phi_k$.
      \item[] H4: There exists some constants $C_5,C_6>0$ such that $C_5 \leq N^{-1} n \omega_k \leq C_6$ for any $k \in U$.
      \item[] H5: There exists some constants $C_7,C_8>0$ such that $C_7 \leq v_k \leq C_8$ for any $k \in U$. There exists some constant $C_9$ such that $\|z_k\| \leq C_9$ for any $k \in U$. Also, the matrix
        \begin{eqnarray} \label{mat:G}
        G & = & \frac{1}{N} \sum_{k \in U} \pi_k \phi_k \omega_k v_k^{-1} z_k z_k^{\top}
        \end{eqnarray}
      is invertible, and the constant $a$ chosen is such that $\|G^{-1}\| \leq a^{-1}$.
    \end{itemize}

\noindent Assumptions (H1) and (H2) are related to the sampling design. In case of sampling with equal probabilities, we have $\pi_k=n/N$ and Assumption (H1) is automatically fulfilled with $C_1=C_2=1$. The assumption (H1) means that in case of sampling with unequal probabilities, the inclusion probabilities do not depart much from that obtained when sampling with equal probabilities. In Assumption (H2), the quantity $\sup_{k \neq l \in U} \left(n \left|1-\frac{\pi_{kl}}{\pi_k\pi_l}\right|\right)$ is a measure of dependency in the selection of units. This quantity is equal to zero when the units in the population are selected independently, which is known as Poisson sampling (Fuller, 2009, p. 13). This assumption is satisfied for many sampling designs like stratified simple random sampling or rejective sampling, see for example Cardot et al. (2013). Both assumptions (H1) and (H2) are classical in survey sampling. \\

\noindent Assumption (H3) is related to the response mechanism. It is assumed that the response probabilities are bounded below from zero, i.e. that all units in the population have a strictly positive probability to answer the survey. Assumption (H4) is related to the imputation mechanism. If $\omega_k=N/n$ for any unit $k \in U$, all the responding units have the same probability of being selected to fill-in a missing value. It is thus assumed in (H4) that when selecting the random residuals, no extreme imputation weight will dominate the others. The assumption (H5) is in particular related to the choice of the regularizing parameter $a$, and is needed to guarantee the point-wise convergence of the estimator $\hat{B}_{ar}$ for the regression coefficient. A similar assumption is considered in Cardot et al. (2013).

\begin{prop} \label{prop3}
  Assume that the imputation model (\ref{imp:mod}) holds, and that assumptions (H1)-(H5) hold. Then:
    \begin{eqnarray}
      E \left\{\|\hat{B}_{ar}-\beta\|^2 \right\} & = & O(n^{-1}). \label{prop3:eq1}
    \end{eqnarray}
\end{prop}

\section{Balanced random imputation} \label{sec:bal:imput:model}

\subsection{Motivation} \label{ssec:41}

\noindent In practice, a survey serves multiple purposes. On the one hand, the survey designer is interested in estimating aggregate parameters such as totals, and the variance of total estimates needs to be kept as small as possible. On the other hand, the survey data are used by secondary analysts who are interested in other parameters of interest which may be related to the distribution of the imputed variable, like quantiles. Therefore, a random imputation method is used to fill-in the missing values in order to preserve the distribution of the imputed variable. In the same time, the random residuals need to be generated in such a way that the variance of $\hat{t}_{yI}$ does not suffer from the variance imputation. \\

\noindent The drawback of random regression imputation lies indeed in an additional variability for the estimation of $t_y$, called the imputation variance. The imputed estimator of the total may be written as
    \begin{eqnarray} \label{est:ht:imp:2}
      \hat{t}_{yI} & = & \sum_{k \in S} d_k r_k y_k + \sum_{k \in S} d_k (1-r_k) (z_k^{\top} \hat B_{ar}) + \sum_{k \in S} d_k (1-r_k) (v_k^{1/2} \epsilon_k^*).
    \end{eqnarray}
The imputation variance is due to the third term on the right-hand side only. This imputation variance is completely eliminated if the random residuals are selected so that
    \begin{eqnarray} \label{bal:eq}
      \sum_{k \in S} d_k (1-r_k) v_k^{1/2} \epsilon_k^*
      & = & E_I\left\{\sum_{k \in S} d_k (1-r_k) v_k^{1/2} \epsilon_k^*\right\} \nonumber \\
      & = & \sum_{k \in S} \left\{d_k (1-r_k) v_k^{1/2} \right\} \bar{e}_r
    \end{eqnarray}
with $\bar{e}_r=\sum_{j \in S} \tilde{\omega}_j r_j e_j$. \\

\subsection{Exact balanced random imputation} \label{ssec:42}

\noindent Our goal is therefore to generate the random residuals in such a way that equation (\ref{bal:eq}) holds. A natural idea would be to select the random residuals $\epsilon_k^*$ directly with replacement from the set $E_r$ of observed residuals. Unfortunately, to the best of our knowledge, there does not exist any general with-replacement sampling design which enables to select the random residuals such that equation (\ref{bal:eq}) holds. In order to be able to use the cube method presented in Section \ref{ssec:13}, we follow the approach in Chauvet et al. (2011) and proceed in 4 steps:
\begin{enumerate}
  \item We build a population $U^*$ of $n_m \times n_r$ cells, each row being associated to one non-respondent and each column being associated to one respondent.
  \item To each cell $(k,l) \in U^*$, we associate a selection probability $\psi_{kl}=\tilde{\omega_l}$ (see equation \ref{pr:epsk:star}) and a value $x_{kl}^0=d_k v_k^{1/2} \psi_{kl} e_l$.
  \item We apply the flight phase of the cube method on population $U^*$, with inclusion probabilities $\psi_{U^*}=(\psi_{11},\ldots,\psi_{n_m n_r})^{\top}$, by balancing on the variable $x_{kl}^0$. From equation (\ref{bal:eq:2}), we obtain at the end of the flight phase a random vector $\tilde{I}_{U^*}=(\tilde{I}_{11},\ldots,\tilde{I}_{n_m n_r})^{\top}$
      such that
     \begin{eqnarray} \label{bal:eq:3}
      \sum_{(k,l) \in U^*} \frac{x_{kl}^0}{\psi_{kl}} \tilde{I}_{kl} & = & \sum_{(k,l) \in U^*} x_{kl}^0.
     \end{eqnarray}
  \item For any non-respondent $k$, the imputed residual is
    \begin{eqnarray} \label{cdh:imp:value}
      \epsilon_k^{*} & = & \sum_{l \in S_r} \tilde{I}_{kl} e_l.
    \end{eqnarray}
\end{enumerate}

\noindent From equations (\ref{bal:eq:3}) and (\ref{cdh:imp:value}), it is easily shown that equation (\ref{bal:eq}) holds. Therefore, the imputation variance of $\hat{t}_{yI}$ is completely eliminated under the proposed imputation procedure. This is an advantage as compared to the balanced imputation procedure in Chauvet et al. (2011), where the balancing constraint (\ref{bal:eq:3}) was not exactly respected, and the imputation variance was therefore not fully eliminated. A drawback of the proposed method is that a missing residual is not necessarily replaced by an observed estimated residual, but may be replaced by a weighted mean of observed estimated residuals, which may result in a bias in the estimation of the distribution function. However, by adding $n_m$ balancing variables in the imputation procedure, we ensure that $\epsilon_k^*$ is an observed residual for at least $n_m-1$ units. The additional $n_m$-vector of balancing variables is
    \begin{eqnarray} \label{add:bal}
      x & = & (x^1,\ldots,x^i,\ldots,x^{n_m})^{\top},
    \end{eqnarray}
with $x_{kl}^{i}=\psi_{kl} 1(k=i)$ for the cell $(k,l)$, see Chauvet et al. (2011). We prove in Proposition \ref{prop5} that using this additional set of balancing variables in the imputation process enables to preserve the distribution of the imputed variable.

\subsection{An illustration of the proposed method} \label{ssec:illust}

\noindent We illustrate the proposed imputation method on a small sample, based on an example presented in Thompson (2002, p. 70). In a population $U$ of $N=53$ persons in a lecture theater, each person $k$ is asked to write down a guess of the amount of money (variable $z_{1k}$) he/she is carrying. A simple random sample of $n=10$ persons is then selected, and each person is asked to write down the exact amount of money (variable $y_k$) he/she is carrying. For this illustration, we consider that the variable $y_k$ is missing for $4$ persons in the sample. The data are presented in Table \ref{tab:ex:thompson}. \\

\begin{table}
\def~{\hphantom{0}}
\centering
\caption{Values of the guess of the amount of money, of the true amount of money and of observed residuals for a simple random sample of $n=10$ persons}{
\begin{tabular}{|c||c|c|c|c|c|c|c|c|c|c|}
  \hline
  Unit $k$ & 1    & 2    & 3     & 4    & 5    & 6     & 7    & 8   & 9 & 10 \\ \hline
  $z_{1k}$ & 8.35 & 1.5  & 10    & 0.6  & 7.5  & 7.95  & 0.95 & 4.4 & 1 & 0.5 \\
  $y_k$    & 8.75 & 2.55 & 9     & 1.1	& 7.5  & 5     &      &     &   &     \\
  $e_k$    & 0.30 & 0.93 & -0.14 & 0.69	& 0.15 & -0.89 &      &     &   &     \\
  \hline
\end{tabular}}
\label{tab:ex:thompson}
\end{table}

\noindent We consider so-called ratio imputation, which is obtained from (\ref{val:imp}) in case of a single auxiliary variable ($z_k=z_{1k}$) and with $v_k=z_{1k}$. This leads to the ratio imputation model 
    \begin{equation} \label{imp:mod:3}
      m:~y_k=\beta z_{1k}+z_{1k}^{1/2} \epsilon_k,
    \end{equation}
which is currently used in business surveys. We use equal imputation weights $\omega_k=1$. This leads to
    \begin{eqnarray} \label{val:imp:rimp}
      y_k^*=\hat{B}_{r} z_{1k} + z_{1k}^{1/2} \epsilon_k^* & \textrm{ where } & \hat{B}_{r} = \frac{\sum_{k \in S} r_k y_k}{\sum_{k \in S} r_k z_{1k}}.
    \end{eqnarray}
In this example, we obtain $\hat{B}_{r}=0.94$. The observed residuals $e_k$ for respondents are given in the last line of Table \ref{tab:ex:thompson}. We have
	\begin{eqnarray} \label{ex:bal:eq:1}
    \sum_{k \in S} d_k (1-r_k) v_k^{1/2} \bar{e}_r & = & 4.38.
    \end{eqnarray}

\noindent In order to impute the missing values, we create a table of $4 \times 6$ cells with one row for each non-respondent and one column for each observed residual. We then draw a sample of $4$ cells by means of the flight phase of the cube method. The result is presented in Table \ref{tab:ex:thompson:2}. The $4$-th cell on row $1$ is selected, which means that we take $\epsilon_{7}^{\star}=e_4=0.69$. Similarly, the $1$-th cell on row $3$ is selected so that $\epsilon_9^{\star}=e_1$, and the $1$-th cell on row $4$ is selected so that we take $\epsilon_{10}^{\star}=e_1$. On row $2$, we obtain $\tilde{I}_{21}=0.61$ and $\tilde{I}_{26}^{\star}=0.39$, so that we take $\epsilon_9^{\star}=0.61*e_1+0.39*e_6=-0.17$. With these imputed residuals, we obtain
	\begin{eqnarray} \label{ex:bal:eq:2}
    \sum_{k \in S} d_k (1-r_k) v_k^{1/2} \epsilon_k^* & = & 4.38
	\end{eqnarray}
so that from equation (\ref{ex:bal:eq:1}), the balancing equation is exactly respected. We present in Table \ref{tab:ex:thompson:3} another possible set of imputed residuals. It can be shown that equation (\ref{ex:bal:eq:1}) holds, so that the balancing equation is satisfied and the imputation variance for the total is eliminated.

\begin{table}
\def~{\hphantom{0}}
\centering
\caption{A first example of balanced random imputation}{
\begin{tabular}{|c||c|c|c|c|c|c||c|c|}
  \hline
  Non-respondent     & \multicolumn{6}{|c||}{Observed residuals}    & Imputed & Imputed \\ 
  $k$                & 0.30 & 0.93 & -0.14 & 0.69	& 0.15 & -0.89  & residual $\epsilon_k^*$ & value $y_k^*$ \\ \hline
  $7$                & 0    & 0    & 0     & 1      & 0    & 0      &  0.69  & 1.57 \\
  $8$                & 0.61 & 0    & 0     & 0      & 0    & 0.39   &  -0.17 & 3.80 \\
  $9$                & 1    & 0    & 0     & 0      & 0    & 0      &  0.30  & 1.24 \\
  $10$               & 1    & 0    & 0     & 0      & 0    & 0      &  0.30  & 0.68 \\
  \hline
\end{tabular}}
\label{tab:ex:thompson:2}
\end{table}

\begin{table}
\def~{\hphantom{0}}
\centering
\caption{A second example of balanced random imputation}{
\begin{tabular}{|c||c|c|c|c|c|c||c|c|}
  \hline
  Non-respondent     & \multicolumn{6}{|c||}{Observed residuals}    & Imputed & Imputed \\ 
  $k$                & 0.30 & 0.93 & -0.14 & 0.69	& 0.15 & -0.89  & residual $\epsilon_k^*$ & value $y_k^*$ \\ \hline
  $7$                & 0    & 0.83 & 0     & 0      & 0    & 0.17   &  0.62   & 1.50 \\
  $8$                & 1    & 0    & 0     & 0      & 0    & 0      &  0.30   & 4.78 \\
  $9$                & 0    & 0    & 0     & 0      & 0    & 1      &  -0.89  & 0.06 \\
  $10$               & 0    & 0    & 0     & 1      & 0    & 0      &  0.69   & 0.96 \\
  \hline
\end{tabular}}
\label{tab:ex:thompson:3}
\end{table}

\subsection{Properties of balanced random imputation} \label{ssec:prop}

It is shown in Proposition \ref{prop4} below that the imputed estimator of the total is mean-square consistent for the true total. Also, we prove in Proposition \ref{prop5} that the imputed distribution function under the proposed exact balanced imputation procedure is consistent for the population distribution function.

\begin{prop} \label{prop4}
  Assume that the imputation model (\ref{imp:mod}) holds, and that assumptions (H1)-(H5) hold. Assume that the exact balanced imputation procedure is used. Then:
    \begin{eqnarray}
      E[\left\{N^{-1}(\hat{t}_{yI}-t_y)\right\}^2] & = & O(n^{-1}). \label{prop4:eq1}
    \end{eqnarray}
\end{prop}

\begin{prop} \label{prop5}
  We assume that assumptions (H1)-(H5) hold. Assume that $F_{\epsilon}$ is absolutely continuous. Assume that the exact balanced imputation procedure is used, and that the balancing variables in (\ref{add:bal}) are added. Then:
    \begin{eqnarray}
      E\left|\hat{F}_I(t)-F_N(t)\right| & = & o(1). \label{prop5:eq1}
    \end{eqnarray}
\end{prop}

%
%

\section{Simulation study} \label{sec:simu:study}

\noindent We conducted a simulation study to test the performance of several imputation methods in terms of relative bias and relative efficiency. We first generated $2$ finite populations of size $N=10,000$, each containing one study variable $y$ and one auxiliary variable $z_1$. In each population, the variable $z_1$ was first generated from a Gamma distribution with shape and scale parameters equal to $2$ and $5$, respectively. Then, given the $z_1$-values, the $y$-values were generated according to the model $y_k=\beta~z_{1k}+ z_{1k}^{1/2}~\epsilon_k$ presented in equation (\ref{imp:mod:3}). The parameter $\beta$ was set to $1$ and the $\epsilon_k$ were generated according to a normal distribution with mean $0$ and variance $\sigma^2$, whose value was chosen to lead to a coefficient of determination ($R^2$) approximately equal to $0.36$ for population 1 and $0.64$ for population 2. \\

\noindent We were interested in estimating two parameters: the population total of the $y$-values, $t_y$ and the finite population distribution function, $F_{N}(t)$ for $t=t_{\alpha}$, where $t_{\alpha}$ is the $\alpha$-th population quantile.  We considered $\alpha=0.25$ and $0.50$ in the simulation. From each population, we selected $1,000$ samples of size $n=100$ by means of rejective sampling also called conditional Poisson sampling (e.g., Hajek, 1964) with inclusion probabilities, $\pi_k$, proportional to $z_k$. That is, we have $\pi_k=nz_k/t_z,$ where $t_z=\sum_{k \in U} z_k$. Then, in each generated sample, nonresponse to item $y$ was generated according to two nonresponse mechanisms which are described below:
\begin{itemize}
  \item[MCAR:] uniform response mechanism, where all the units in $U$ have the same probability of response $\phi_0$. We used $\phi_0=0.5$ and $\phi_0=0.75$.
  \item[MAR:] the probability $\phi_k$ of response attached to unit $k$ is defined as
    \begin{eqnarray} \label{eq:mar}
      \log\left(\frac{\phi_k}{1-\phi_k}\right) & = & \lambda_0+\lambda_1 z_{1k},
    \end{eqnarray}
  where the parameters $\lambda_0$ and $\lambda_1$ were chosen so that the average $\bar{\phi}$ of the $\phi_k$'s was approximately equal to 0.5, or approximately equal to 0.75.
\end{itemize}

\noindent In each sample containing respondents and nonrespondents, imputation was performed according to three methods, all motivated by the imputation model (\ref{imp:mod:3}). The imputed values are given by
    \begin{eqnarray} \label{imp:mec:rat}
      y_k^* & = & \hat{B}_{r} z_{1k} + z_{1k}^{1/2} \epsilon_k^*.
    \end{eqnarray}
For deterministic ratio imputation (DRI), the imputed values are given by (\ref{imp:mec:rat}) with $\epsilon_k^*=0$ for all $k$. The imputed values for random ratio imputation (RRI) are given by (\ref{imp:mec:rat}), where the residuals $\epsilon_k^*$ are selected independently and with replacement. The imputed values for exact balanced ratio imputation (EBRI) are given by (\ref{imp:mec:rat}) where the residuals $\epsilon_k^*$ are selected so that the balancing constraint (\ref{bal:eq}) is exactly satisfied. \\

\noindent Then, we computed the imputed estimator of $t_y$ given by (\ref{est:ht:imp}), and the imputed estimator of $F_N(t)$ given by (\ref{est:ht:df:imp}). As a measure of the bias of an estimator $\hat{\theta}_I$ of a parameter $\theta$, we used the Monte Carlo percent relative bias
    \begin{equation} \label{rb}
      \mbox{RB}(\hat{\theta}_I)=\frac{E_{MC}(\hat{\theta}_I)-\theta}{\theta}\times 100,
    \end{equation}
where $E_{MC}(\hat{\theta}_I)=\sum_{r=1}^{1000} \hat{\theta}^{(r)}_I/1000,$ and $\hat{\theta}^{(r)}_I$ denotes the estimator $\hat{\theta}_I$ in the $r$-th sample, $r=1,~\ldots~,1000$. As a measure of variability of $\hat{\theta}_I$, we used the Monte Carlo mean square error
    \begin{equation} \label{mse}
      \mbox{MSE}(\hat{\theta}_I)=\frac{1}{1000} \sum_{r=1}^{1000} (\hat{\theta}^{(r)}_I-\theta)^2.
    \end{equation}
Let $\hat{\theta}^{(DRI)}_I$, $\hat{\theta}^{(RRI)}_I$, and $\hat{\theta}^{(EBRI)}_I$ denote the estimator $\hat{\theta}_I$ under deterministic ratio imputation, random ratio imputation and exact balanced ratio imputation, respectively. In order to compare the relative efficiency of the imputed estimators, using $\hat{\theta}^{(RRI)}_I$ as the reference, we used
    \begin{equation} \label{re}
      \mbox{RE}=\frac{\mbox{MSE}(\hat{\theta}^{(.)}_I)}{\mbox{MSE}(\hat{\theta}^{(RRI)}_I)}.
    \end{equation}
Monte Carlo measures for $\hat{F}_I(t)$ were obtained from (\ref{rb})-(\ref{re}) by replacing $\hat{\theta}_I$ with $\hat{F}_{I}(t)$ and $\theta_N$ with $F_{N}(t)$. \\

\begin{table}
\def~{\hphantom{0}}
\centering
\caption{Monte Carlo percent relative bias of the imputed estimator and relative efficiency}{
\begin{tabular}{|rl|ccc|ccc|}
\hline\hline
                    & & DRI & RRI & EBRI & DRI & RRI & EBRI \\\hline \hline
                    & & \multicolumn{6}{|c|}{MCAR} \\ \cline{3-8}
                    & & \multicolumn{3}{|c|}{$\phi_0=0.5$} & \multicolumn{3}{|c|}{$\phi_0=0.75$} \\ \hline
{\tt Population 1}  & RB & 0.47 & 0.50  & 0.47 & 0.30 & 0.33 & 0.30 \\
                    & RE & 0.79 & 1     & 0.79 & 0.79 & 1    & 0.79 \\ \hline
{\tt Population 2}  & RB & 0.17 & 0.26  & 0.17 & 0.16 & 0.25 & 0.16 \\
                    & RE & 0.79 & 1     & 0.79 & 0.79 & 1    & 0.79 \\ \hline \hline
                    & & \multicolumn{6}{|c|}{MAR} \\ \cline{3-8}
                    & & \multicolumn{3}{|c|}{$\bar{\phi}=0.5$} & \multicolumn{3}{|c|}{$\bar{\phi}=0.75$} \\ \hline
{\tt Population 1}  & RB & 0.28 & 0.30  & 0.28 & 0.45  & 0.62  & 0.45 \\
                    & RE & 0.69 & 1     & 0.69 & 0.72  & 1     & 0.72 \\ \hline
{\tt Population 2}  & RB & 0.02 & -0.06 & 0.02 & -0.18 & -0.14 & -0.18 \\
                    & RE & 0.70 & 1     & 0.70 & 0.74  & 1     & 0.74 \\ \hline \hline
\end{tabular}}
\label{tab:rb:re}
\end{table}

\noindent Table \ref{tab:rb:re} shows the values of relative bias and relative efficiency corresponding to the imputed estimator $\hat{t}_{yI}$. It is clear from Table \ref{tab:rb:re} that $\hat{t}_{yI}$ was approximately unbiased in all the scenarios, as expected. In terms of relative efficiency, results showed that DRI and EBRI lead to the smallest mean square error for the estimation of a total. This result is not surprising since the imputation variance is identically equal to zero for both imputation methods. We note that DRI and EBRI were particulary efficient in the MAR case. \\

\renewcommand{\arraystretch}{0.6}
\begin{table}
\def~{\hphantom{0}}
\centering
\caption{Monte Carlo percent relative bias of the imputed estimator
    of the distribution function and relative efficiency}{
\begin{tabular}{|rlll|ccc|ccc|}
\hline\hline
                    & & & & DRI & RRI & EBRI & DRI & RRI & EBRI \\ \hline \hline
                    & &          & & \multicolumn{6}{|c|}{MCAR} \\ \cline{5-10}
                    & & $\alpha$ & & \multicolumn{3}{|c|}{$\phi_0=0.5$} & \multicolumn{3}{|c|}{$\phi_0=0.75$} \\ \hline
{\tt Population 1}  & & 0.25 & RB & -41.3 & -1.6 & -2.7 & -31.3 & -1.1 & -2.0 \\
                    & &      & RE & 2.03  & 1    & 0.94 & 1.66  & 1    & 0.94 \\
                    & & 0.50 & RB & -4.7  & -1.3 & -0.9 & -3.6  & -0.7 & -0.6 \\
                    & &      & RE & 1.22  & 1    & 0.98 & 1.13  & 1    & 0.97 \\ \hline
{\tt Population 2}  & & 0.25 & RB & -26.7 & -0.7 & -1.4 & -22.2 & -0.5 & -1.1 \\
                    & &      & RE & 1.45  & 1    & 0.93 & 1.34  & 1    & 0.94 \\
                    & & 0.50 & RB & -2. 7 & -0.3 & -0.1 & -2.1  & -0.1 & 0.1  \\
                    & &      & RE & 1.09  & 1    & 0.97 & 1.07  & 1    & 0.97 \\ \hline \hline
                    & &          & & \multicolumn{6}{|c|}{MAR} \\ \cline{5-10}
                    & & $\alpha$ & & \multicolumn{3}{|c|}{$\phi_0=0.5$} & \multicolumn{3}{|c|}{$\phi_0=0.75$} \\ \hline
{\tt Population 1}  & & 0.25 & RB & -42.4 & -0.3 & -0.9 & -24.2 & -2.5 & -3.5 \\
                    & &      & RE & 2.11  & 1    & 0.89 & 1.37  & 1    & 0.90 \\
                    & & 0.50 & RB & 2.1   & -0.0 & 0.2  & 5.6   & -1.5 & -0.2 \\
                    & &      & RE & 1.18  & 1    & 0.95 & 1.12  & 1    & 0.96 \\ \hline
{\tt Population 2}  & & 0.25 & RB & -15.4 & 0.6  & 0.4  & -3.4  & 1.3  & 2.0 \\
                    & &      & RE & 1.17  & 1    & 0.93 & 1.00  & 1    & 0.93 \\
                    & & 0.50 & RB & 0.2   & 0.1  & -0.3 & 2.6   & -0.7 & -0.2 \\
                    & &      & RE & 1.09  & 1    & 1.00 & 1.02  & 1    & 0.96 \\ \hline \hline
\end{tabular}}
\label{tab:rb:dist:func}
\end{table}

\noindent We now turn to the distribution function, $F_{N}(t)$. Table \ref{tab:rb:dist:func} shows the relative bias and relative efficiency corresponding to the imputed estimator $\hat{F}_{I}(t)$. As expected, the estimators under deterministic ratio imputation were considerably biased, and the absolute relative bias can be as high as $42.4 \% $. In terms of relative bias, both RRI and EBRI showed almost no bias, except for $t_{0.25}$ in the case of balanced imputation. These results can be explained by the fact that both imputation methods succeeded in preserving the distribution of the study variable $y$. Also, we note that the imputed estimator $\hat{F}_{I}(t)$ under exact balanced ratio imputation was more efficient than the corresponding estimator under random ratio imputation in all the scenarios with a value of relative efficiency varying from $0.89$ to $1.00$. The lower values of RE were obtained in the MAR case.

\section{Final remarks} \label{sec:conc}

\noindent In this paper, we considered estimation under item non-response. We proposed an exact balanced random imputation procedure, where the imputation variance is completely eliminated for the estimation of a total. We also proved that the proposed imputation procedure leads to mean-square consistent estimators for a total and for a distribution function. \\

\noindent We have not considered the problem of variance estimation in the context of the proposed balanced random imputation. Variance estimation for the imputed estimator of the total is fairly straightforward, since the imputed estimator is identical to that under deterministic regression imputation. Variance estimation for the imputed distribution function is currently under investigation. \\

\noindent When studying relationships between study variables, Shao and Wang (2002) proposed a joint random regression imputation procedure which succeeds in preserving the relationship between these variables, and a balanced version of their procedure was proposed by Chauvet and Haziza~(2012). Extending the exact balanced random procedure to this situation is a matter for further research.

\section*{References}

\begin{description}

\item Bhatia, R. (1997).
\newblock{Matrix analysis}.
\newblock {\em Springer-Verlag}.

\item Boistard, H., and Chauvet, G., and Haziza, D. (2016).
\newblock{Doubly robust inference for the distribution function in the presence of missing survey data.}
\newblock{\em Scandinavian Journal of Statistics}, {\bf 43}, 683-699.

\item Cardot, H., and Goga, C., and Lardin, P. (2013).
\newblock{Uniform convergence and asymptotic confidence bands for model-assisted estimators of the mean of sampled functional data.}
\newblock{\em Electronic Journal of Statistics}, {\bf 7}, 562-596.

\item Chauvet, G., and Deville, J.C., and Haziza, D. (2011).
\newblock{On balanced random imputation in surveys.}
\newblock{\em Biometrika}, {\bf 98(2)}, 459-471.

\item Chauvet, G. and Haziza, D. (2012).
\newblock{Fully efficient estimation of coefficients of correlation in the presence of imputed survey data.}
\newblock{\em Canadian Journal of Statistics}, {\bf 40(1)}, 124-149.

\item Chauvet, G. and Till\'e, Y. (2006).
\newblock{A fast algorithm for balanced sampling.}
\newblock{\em Computational Statistics}, {\bf 21(1)}, 53-62.

\item Chen, J., and Rao, J. N. K., and Sitter, R. R. (2000).
\newblock{Efficient random imputation for missing data in complex surveys.}
\newblock{\em Statistica Sinica}, {\bf 10(4)}, 1153-1169.

\item Deville (2006).
\newblock{Random imputation using balanced sampling.}
\newblock{\em Presentation to the Joint Statistical Meeting of the American Statistical Association, Seattle, USA}.

\item Deville, J. C., and S\"arndal, C. E. (1994).
\newblock{Variance estimation for the regression imputed Horvitz-Thompson estimator.}
\newblock{\em Journal of Official Statistics}, {\bf 10}, 381-394.

\item Deville, J. C., and Till\'e, Y. (2004).
\newblock{Efficient balanced sampling: the cube method.}
\newblock{\em Biometrika}, {\bf 91(4)}, 893-912.

\item Fay, R.E. (1996).
\newblock{Alternative paradigms for the analysis of imputed survey data.}
\newblock{\em Journal of the American Statistical Association}, {\bf 91(434)}, 490-498.

\item Fuller, W. A. (2009).
\newblock{Sampling statistics.}
\newblock{\em Wiley}.

\item Fuller, W. A., and Kim, J. K. (2005).
\newblock{Hot deck imputation for the response model.}
\newblock{\em Survey Methodology}, {\bf 31}, 139-149.

\item Gregoire, T.G., and Valentine, H.T. (2008).
\newblock{Sampling strategies for natural resources and the environment.}
\newblock{\em Chapman \& Hall: Boca Raton}.

\item Hajek, J. (1964).
\newblock{Asymptotic theory of rejective sampling with varying probabilities from a finite population.}
\newblock{\em The Annals of Mathematical Statistics}, {\bf 35}, 1491-1523.

\item Hasler, C., and Till\'e, Y. (2014).
\newblock{Fast balanced sampling for highly stratified population.}
\newblock{\em Computational Statistics and Data Analysis}, {\bf 74}, 81-94.

\item Haziza, D. (2009).
\newblock{Imputation and inference in the presence of missing data.}
\newblock{\em Handbook of Statistics}, {\bf 29}, 215-246.

\item Haziza, D., and Nambeu, C., and Chauvet, G. (2016).
\newblock{Doubly robust imputation procedures for finite population means in the presence of a large number of zeros.}
\newblock{\em Canadian Journal of Statistics}, {\bf 42(4)}, 650-669.

\item Isaki, C. T., and Fuller, W. A. (1982).
\newblock{Survey design under the regression superpopulation model.}
\newblock{\em Journal of the American Statistical Association}, {\bf 77(377)}, 89-96.

\item Kalton, G., and Kish, L. (1981).
\newblock{Two efficient random imputation procedures.}
\newblock{\em Proceedings of the Survey Research Methods, American Statistical Association}, 146-151.

\item Kalton, G., and Kish, L. (1984).
\newblock{Some efficient random imputation methods.}
\newblock{\em Communications in Statistics-Theory and Methods}, {\bf 13(16)}, 1919-1939.

\item Kim, J. K., and Fuller, W. (2004).
\newblock{Fractional hot deck imputation.}
\newblock{\em Biometrika}, {\bf 91(3)}, 559-578.

\item Ohlsson, E. (1998).
\newblock{Sequential Poisson Sampling.}
\newblock{\em Journal of Official Statistics}, {\bf 14(2)}, 149-162.

\item Pfeffermann, D. (2009).
\newblock{Inference under informative sampling.}
\newblock{\em Handbook of Statistics}, {\bf 29(B)}, 455-487.

\item Rubin, D. B. (1976).
\newblock{Inference and missing data.}
\newblock{\em Biometrika}, {\bf 63(3)}, 581-592.

\item Rubin, D. B. (1983).
\newblock{Conceptual issues in the presence of nonresponse.}
\newblock{\em Incomplete data in sample surveys}, {\bf 2}, 123-142.

\item S\"arndal, C.-E. (1992).
\newblock{Method for estimating the precision of survey estimates when imputation has been used.}
\newblock{\em Survey Methodology}, {\bf 18}, 241-252.

\item S\"arndal, C. E., and Swensson, B., and Wretman, J. (1992).
\newblock{Model assisted survey sampling.}
\newblock{\em Springer}.

\item Shao, J., and Wang, H. (2002).
\newblock{Sample correlation coefficients based on survey data under regression imputation.}
\newblock{\em Journal of the American Statistical Association}, {\bf 97(458)}, 544-552.

\item Thompson, W. A. (2002).
\newblock{Sampling.}
\newblock{\em Wiley}.

\item Till\'e, Y. (2006).
\newblock{Sampling algorithms.}
\newblock{\em Springer: New-York}.

\end{description}

\appendix

\newpage

\section{Flight phase of the cube method (Till\'e, 2006, p. 160)} \label{app:flight}

\noindent We define the balancing matrix as $A=(x_1/\pi_1,\ldots,x_N/\pi_N)$. We initialize with $\pi_U(0)=\pi_U$. Next, at time $t=0,\ldots,T$, repeat the three following steps. \\

\noindent Step 1: Let $E(t)=F(t) \cap  \mbox{Ker} A,$ where
    \begin{eqnarray*}
      F(t) & = & \{v \in \mathbb{R}^N : v_k=0 \textrm{ if } \pi_{k}(t) \textrm{ is an integer}\},
    \end{eqnarray*}
with $\pi_U(t)=\left(\pi_{1}(t),\ldots,\pi_{N}(t)\right)^{\top}$. If $E(t) \neq \{0\}$, generate any vector $v(t) \neq 0$ in $E(t)$, random or not. \\

\noindent Step 2: Compute the scalars $\lambda_{1}^*(t)$ and $\lambda_{2}^*(t)$, which are the largest values of $\lambda_1(t)$ and $\lambda_2(t)$ such that
    \begin{eqnarray*}
      0 \leq \pi_U(t)+\lambda_{1}(t) v(t) \leq 1 & \textrm{ and } & 0 \leq \pi_U(t)-\lambda_2(t) v(t) \leq 1,
    \end{eqnarray*}
where the inequalities are interpreted element-wise. Note that $\lambda_1^*(t)>0$ and $\lambda_2^*(t)>0.$ \\

\noindent Step 3: Take $\pi_U(t+1)=\pi_U(t)+\delta_U(t),$ where
    \begin{eqnarray*}
    \delta_U(t) & = & \left\{ \begin{array}{lll}
        \lambda_{1}^*(t) v(t) & \textrm{with probability} & \frac{\lambda_{2}^*(t)}{\lambda_{1}^*(t)+\lambda_{2}^*(t)}, \\
        -\lambda_{2}^*(t) v(t) & \textrm{with probability} & \frac{\lambda_{1}^*(t)}{\lambda_{1}^*(t)+\lambda_{2}^*(t)}. \\
    \end{array} \right.
    \end{eqnarray*}

\noindent The flight phase ends at time $T$, when it is no longer possible to find a non-null vector in $E(T)$. The random vector obtained at the end of the flight phase is $\tilde{I}_U=\pi(T)$.

\section{Proof of Proposition \ref{prop3}}

\begin{lem} \label{prop3:lem0}
We have
    \begin{eqnarray}
      E_{p} \left[(\hat{t}_{y\pi}-t_y)^2\right] & \leq & \left(\sup_{k \neq l \in U} n \left|1-\frac{\pi_{kl}}{\pi_k\pi_l}\right|\right) \sum_{k \in U} \pi_k \left(\frac{y_k}{\pi_k}-\frac{t_y}{n}\right)^2, \label{prop1:eq1} \\
      E_{p} \left[\{\hat{F}_N(t)-F_N(t)\}^2\right] & \leq & \left(\frac{4}{N^2}\right) \left(\sup_{k \neq l \in U} n \left|1-\frac{\pi_{kl}}{\pi_k\pi_l}\right|\right) \sum_{k \in U} \frac{1}{\pi_k}. \label{prop1:eq2}
    \end{eqnarray}
\end{lem}

\begin{lemproof}
The proof is standard, and is therefore omitted.
\end{lemproof}

\begin{lem} \label{prop3:lem1}
  We have:
    \begin{eqnarray}
      E (\|\hat{G}_r-G\|^2) & = & O(n^{-1}).
    \end{eqnarray}
\end{lem}

\begin{lemproof}
  We note $\|\cdot\|_F$ for the Frobenius norm. Using the fact that the spectral norm is smaller than the Frobenius norm, we have
    \begin{eqnarray} \label{proof:prop3:lem1:eq1}
      E (\|\hat{G}_r-G\|^2) & \leq & E(\|\hat{G}_r-G\|_F^2) \nonumber \\
                            & = & E \left[\frac{1}{N^2} \sum_{k \in U} \sum_{l \in U} \omega_k v_k^{-1} (I_k r_k - \pi_k \phi_k) \omega_l v_l^{-1} (I_l r_l - \pi_l \phi_l) tr(z_k z_k^{\top} z_l z_l^{\top})\right] \nonumber \\
                                  & = & T_3+T_4
    \end{eqnarray}
  with
    \begin{eqnarray}
      T_3 & = & \frac{1}{N^2} \sum_{k \in U} \omega_k^2 v_k^{-2} \pi_k \phi_k (1-\pi_k \phi_k) tr(z_k z_k^{\top} z_k z_k^{\top}), \label{proof:prop3:lem1:eq2} \\
      T_4 & = & \frac{1}{N^2} \sum_{k \neq l \in U} \omega_k v_k^{-1} \omega_l v_l^{-1} (\pi_{kl}-\pi_k \pi_l) \phi_k \phi_l tr(z_k z_k^{\top} z_l z_l^{\top}). \label{proof:prop3:lem1:eq3}
    \end{eqnarray}
  Since $tr(z_k z_k^{\top} z_k z_k^{\top}) = \|z_k\|^4$, we obtain from assumptions (H1), (H4) and (H5)
    \begin{eqnarray} \label{proof:prop3:lem1:eq4}
      T_3 & \leq & \left(\frac{(C_6)^2 C_2}{n(C_7)^2}\right) \left(\frac{1}{N} \sum_{k \in U} \|z_k\|^4\right),
    \end{eqnarray}
  which is $O(n^{-1})$ from assumption (H5). Also, we have
     \begin{eqnarray} \label{proof:prop3:lem1:eq5}
      T_4 & \leq & \frac{1}{N^2} \left( \sup_{k \neq l \in U} \left|1-\frac{\pi_{kl}}{\pi_k\pi_l}\right|\right) \sum_{k \neq l \in U} \omega_k v_k^{-1} \omega_l v_l^{-1} \pi_k \pi_l \phi_k \phi_l tr(z_k z_k^{\top} z_l z_l^{\top}). \end{eqnarray}
  Since
     \begin{eqnarray} \label{proof:prop3:lem1:eq6}
      tr(z_k z_k^{\top} z_l z_l^{\top}) = (z_k^{\top} z_l)^2 \leq \|z_k\|^2 \|z_l\|^2,
     \end{eqnarray}
  we obtain
     \begin{eqnarray} \label{proof:prop3:lem1:eq7}
      T_4 & \leq & \frac{1}{N^2} \left( \sup_{k \neq l \in U} \left|1-\frac{\pi_{kl}}{\pi_k\pi_l}\right|\right) \left(\sum_{k \in U} \omega_k v_k^{-1} \pi_k \phi_k \|z_k\|^2\right)^2 \nonumber \\
          & \leq & \frac{(C_2)^2 (C_6)^2}{n(C_7)^2} \left( \sup_{k \neq l \in U} n \left|1-\frac{\pi_{kl}}{\pi_k\pi_l}\right|\right) \left( \frac{1}{N} \sum_{k \in U} \|z_k\|^2 \right)^2,
     \end{eqnarray}
  which is $O(n^{-1})$ from assumptions (H2) and (H5). This completes the proof of Lemma \ref{prop3:lem1}.
\end{lemproof}

\noindent We now consider the proof of Proposition \ref{prop3}. We can write
    \begin{eqnarray} \label{proof:prop3:eq1}
      \hat{B}_{ar}-\beta & = & T_5+T_6,
    \end{eqnarray}
where
    \begin{eqnarray}
      T_5 & = & \hat{G}_{ar}^{-1} \left\{\frac{1}{N} \sum_{k \in S} r_k \omega_k v_k^{-1} z_k (y_k-z_k^{\top} \beta) \right\}, \label{proof:prop3:eq2} \\
      T_6 & = & \hat{G}_{ar}^{-1} \left\{(\hat{G}_{r}-\hat{G}_{ar}) 1(\hat{G}_{ar} \neq \hat{G}_{r}) \right\} \beta. \label{proof:prop3:eq3}
    \end{eqnarray}
We first consider the term $T_5$. We have:
    \begin{eqnarray} \label{proof:prop3:eq4}
      \|T_5\|^2 & \leq & \|\hat{G}_{ar}^{-1}\|^2 \times \left\|\frac{1}{N} \sum_{k \in S} r_k \omega_k v_k^{-1} z_k (y_k-z_k^{\top} \beta) \right\|^2 \\
                & \leq & a^{-2} \times \frac{1}{N^2} \sum_{k,l \in S} r_k r_l \omega_k \omega_l v_k^{-1} v_l^{-1} z_k^{\top} z_l (y_k-z_k^{\top} \beta) (y_l-z_l^{\top} \beta), \nonumber
    \end{eqnarray}
where the second line in (\ref{proof:prop3:eq4}) follows from (\ref{norm:Gar:inv}). Since the sampling design is non-informative and the response mechanism is unconfounded, we can write
    \begin{eqnarray} \label{proof:prop3:eq5}
      E(\|T_5\|^2) & = & E_{pq} E_m (\|T_5\|^2),
    \end{eqnarray}
where $E_{pq}(\cdot)$ stands for the expectation with respect to the sampling design and the response mechanism, and $E_{m}(\cdot)$ stands for the expectation with respect to the imputation model conditionally on $I_U$ and $r_U$. From (\ref{proof:prop3:eq4}), (\ref{proof:prop3:eq5}), and from the assumptions on the imputation model (\ref{imp:mod}), we obtain
    \begin{eqnarray} \label{proof:prop3:eq6}
      E(\|T_5\|^2) & \leq & E_{pq} \left\{ \sigma^2 a^{-2} \times \frac{1}{N^2} \sum_{k \in S} r_k \omega_k^2 v_k^{-1} z_k^{\top} z_k \right\} \nonumber\\
                   & \leq & \left(\frac{\sigma^2 a^{-2} (C_6)^2 C_2}{n C_7}\right) \left(\frac{1}{N} \sum_{k \in U} \|z_k\|^2 \right)
    \end{eqnarray}
where the second line in (\ref{proof:prop3:eq6}) follows from assumptions (H1), (H4) and (H5). From assumption (H5), this leads to $E(\|T_5\|^2)=O(n^{-1})$. \\

\noindent We now consider the term $T_6$, by following the same lines as in Lemma A.1 of Cardot et al.~(2013). We have:
    \begin{eqnarray} \label{proof:prop3:eq7}
      \|T_6\|^2 & \leq & \|\hat{G}_{ar}^{-1}\|^2 \times \left\|(\hat{G}_{r}-\hat{G}_{ar}) 1(\hat{G}_{ar} \neq \hat{G}_{r})\right\|^2 \times \|\beta\|^2 \nonumber \\
                & \leq & a^{-2} \|\beta\|^2 \times \left\|(\hat{G}_{r}-\hat{G}_{ar}) 1(\hat{G}_{ar} \neq \hat{G}_{r})\right\|^2.
    \end{eqnarray}
Since $\|\hat{G}_{ar}-\hat{G}_r\|^2 \leq a^2$, we obtain
    \begin{eqnarray} \label{proof:prop3:eq7}
      E(\|T_6\|^2) & \leq & \|\beta\|^2 \times Pr(\hat{G}_{ar} \neq \hat{G}_{r}).
    \end{eqnarray}
We write
    \begin{eqnarray} \label{proof:prop3:eq8}
      G = \sum_{j=1}^p \eta_{j} u_{j} u_{j}^{\top},
    \end{eqnarray}
where $\eta_{1} \geq \ldots \geq \eta_{p}$ are the non-negative eigenvalues of $G$, with $u_{1},\ldots,u_{p}$ the corresponding orthonormal eigenvectors. We have
    \begin{eqnarray}
      Pr(\hat{G}_{ar} \neq \hat{G}_{r}) & = & Pr(\eta_{pr} \neq a) \nonumber \\
                                        & \leq & Pr\left(|\eta_{pr}-\eta_p| \geq \frac{|\eta_p-a|}{2}\right) \nonumber \\
                                        & \leq & \frac{4}{(\eta_p-a)^2} E(|\eta_{pr}-\eta_p|^2) \label{proof:prop3:eq9} \\
                                        & \leq & \frac{4}{(\eta_p-a)^2} E \|\hat{G}_r-G\|^2 \label{proof:prop3:eq10}
    \end{eqnarray}
where equation (\ref{proof:prop3:eq9}) follows from the Chebyshev inequality, and equation (\ref{proof:prop3:eq10}) follows from the fact that the eigenvalue map is Lipschitzian for symmetric matrices (see Bhatia (1997), chapter 3, and Cardot et al. (2013), p. 580). From (\ref{proof:prop3:eq7}) and (\ref{proof:prop3:eq10}), and using Lemma \ref{prop3:lem1}, we obtain $E(\|T_6\|^2)=O(n^{-1})$. This completes the proof.

\section{Proof of Proposition \ref{prop4}}

\noindent From equation (\ref{prop1:eq1}), we obtain under Assumptions (H1), (H2), (H5) and under the model assumptions that
    \begin{eqnarray}
      E[\left\{N^{-1}(\hat{t}_{y\pi}-t_y)\right\}^2] & = & O(n^{-1}). \label{proof:prop4:eq1}
    \end{eqnarray}
It is therefore sufficient to prove that
    \begin{eqnarray}
      E[\left\{N^{-1}(\hat{t}_{yI}-\hat{t}_{y\pi})\right\}^2] & = & O(n^{-1}). \label{proof:prop4:eq2}
    \end{eqnarray}
We can write $N^{-1}(\hat{t}_{yI}-\hat{t}_{y\pi})=T_7-T_8$, where
    \begin{eqnarray}
      T_7 & = & N^{-1} \sum_{k \in S} d_k (1-r_k) z_k^{\top} (\hat{B}_{ar}-\beta), \label{proof:prop4:eq3} \\
      T_8 & = & \sigma N^{-1} \sum_{k \in S} d_k (1-r_k) v_k^{1/2} \epsilon_k. \label{proof:prop4:eq4}
    \end{eqnarray}
We have
    \begin{eqnarray}
      |T_7|^2 & \leq & N^{-2} \left\| \sum_{k \in S} d_k (1-r_k) z_k \right\|^{2} \times \left\|\hat{B}_{ar}-\beta \right\|^2 \nonumber \\
              & \leq & (C_9/C_1)^2 \left\|\hat{B}_{ar}-\beta \right\|^2, \label{proof:prop4:eq5}
    \end{eqnarray}
where the second line in (\ref{proof:prop4:eq5}) follows from Assuptions (H1) and (H5). From Proposition \ref{prop3}, we obtain $E(|T_7|^2)=O(n^{-1})$. \\

\noindent We now turn to $T_8$. We have $E_m(T_8)=0$, so that
    \begin{eqnarray} \label{proof:prop4:eq6}
      E(T_8^2)=V(T_8)=E_pE_qV_m(T_8).
    \end{eqnarray}
Also, we have
    \begin{eqnarray} \label{proof:prop4:eq7}
      V_m(T_8) & = & \sigma^2 N^{-2} \sum_{k \in S} d_k^2 (1-r_k) v_k,
    \end{eqnarray}
which is $O(n^{-1})$ from Assumptions (H1) and (H5). This completes the proof.

\section{Proof of Proposition \ref{prop5}}

\noindent We can write
    \begin{eqnarray} \label{proof:prop5:eq1}
      \hat{F}_I(t)-F_N(t) & = & \left\{\hat{F}_I(t)-\tilde{F}_I(t)\right\}+\left\{\tilde{F}_I-F_N(t)\right\},
    \end{eqnarray}
where
    \begin{eqnarray} \label{proof:prop5:eq2}
      \tilde{F}_I(t) & = & \hat{N}^{-1} \left\{\sum_{k \in S} d_k r_k 1(y_k \leq t)+\sum_{k \in S} d_k r_k 1(y_k^{**} \leq t) \right\},
    \end{eqnarray}
and $y_k^{**}=z_k^{\top} \hat B_{ar} + v_k^{1/2} \epsilon_k^{**}$ is the imputed value under the balanced random imputation procedure of Chauvet et al. (2011). \\

\noindent Since the number of units such that $0<\tilde{I}_{kl}<1$ at the end of the flight phase is bounded, we have $y_k^*=y_k^{**}$ for all units in $S_m$ but a bounded number of units. Therefore, there exists some constant $C$ such that
    \begin{eqnarray} \label{proof:prop5:eq3}
      |\hat{F}_I(t)-\tilde{F}_I(t)| & \leq & \hat{N}^{-1} \times C \sup_{k \in S} d_k \nonumber \\
                                    & \leq & \frac{C~C_2}{C_1~n},
    \end{eqnarray}
where the second line in (\ref{proof:prop5:eq3}) follows from Assumption (H1). Therefore,
    \begin{eqnarray} \label{proof:prop5:eq4}
      E|\hat{F}_I(t)-\tilde{F}_I(t)| & = & O(n^{-1}).
    \end{eqnarray}
It follows from the proof of Theorem 2 in Chauvet et al. (2011) that
    \begin{eqnarray} \label{proof:prop5:eq5}
      E|\tilde{F}_I(t)-F_N(t)| & = & o(1).
    \end{eqnarray}
From (\ref{proof:prop5:eq1}), the proof is complete.

\end{document}